\documentstyle[12pt]{article}
\input psfig
 1
\def\as{\alpha_{\rm S}}

\def\citenum#1{{\def\@cite##1##2{##1}\cite{#1}}}
\def\citea#1{\@cite{#1}{}}

\def\real{\mathop{\rm Re}}

\def\as{\alpha_{\rm S}}

\def\D{\Delta}

\def\la{\lambda}

\def\m{\mu}

\def\o{\omega}

\def\p{\phi}

\def\ra{\rightarrow}

\def\s{\sigma}

\def\ti{\tilde}

\def\({\left(}
\def\){\right)}

\def\citenum#1{{\def\@cite##1##2{##1}\cite{#1}}}
\def\citea#1{\@cite{#1}{}}

\def\l1vt{\vec{l_{1\perp}}}

\def\rt{r_{\perp}}
\def\bt{b_{\perp}}
\def\rt2{r^2_{\perp}}

\def\bt2{$b^2_t$}

\def\jol1{$J_0(\,l_{1\perp}\,r_{\perp}\,)$}

\def\citea#1{\@cite{#1}{}}

\def\VEV#1{\left\langle #1\right\rangle}

\def\beq{\begin{equation}}
\def\eeq{\end{equation}}
\def\bea{\begin{eqnarray}}
\def\eea{\end{eqnarray}}

\def\eq#1{{Eq.~(\ref{#1})}}

\def\bbbz{{\mathchoice {\hbox{$\sf\textstyle Z\kern-0.4em Z$}}
{\hbox{$\sf\textstyle Z\kern-0.4em Z$}}
{\hbox{$\sf\scriptstyle Z\kern-0.3em Z$}}
{\hbox{$\sf\scriptscriptstyle Z\kern-0.2em Z$}}}}
\def\npb#1#2#3{    {\it Nucl. Phys. }{\bf B#1} (19#2) #3}
\def\plb#1#2#3{    {\it Phys. Lett. }{\bf B#1} (19#2) #3}
\def\prd#1#2#3{    {\it Phys. Rev. }{\bf D#1} (19#2) #3}

\def\zpc#1#2#3{    {\it Z. Phys. }{\bf C#1} (19#2) #3}

\def\sjnp#1#2#3{   {\it Sov. J. Nucl. Phys. }{\bf #1} (19#2) #3}

\relax
\begin{document}
\begin{titlepage}
\noindent
 August  1995   \hfill  CBPF-NF-061/95 \,\,\,\,{\bf hep\,-\,ph\,/\,9508414}
\\[4ex]
\begin{center}
{\Large\bf LANDAU - POMERANCHUK - MIGDAL}\\[1.4ex]
{\Large \bf EFFECT  FOR NUCLEAR MATTER}\\[1.4ex]
{\Large \bf IN QCD} \\[11ex]

{\large E U G E N E \,\, L E V I N}   $^{*)} $
\footnotetext{$^{*}$ Email: levin@lafex.cbpf.br} \\[1.5ex]
{\it  LAFEX, Centro Brasileiro de Pesquisas F\'\i sicas  (CNPq)}\\
{\it Rua Dr. Xavier Sigaud 150, 22290 - 180 Rio de Janeiro, RJ, BRASIL}
\\{\it and}\\
{\it Theory Department, Petersburg Nuclear Physics Institute}\\
{\it 188350, Gatchina, St. Petersburg, RUSSIA}\\[2ex]
\end{center}
\vspace{3cm}
{\large \bf Abstract:}
Soft photon and gluon radiation off a fast quark propagating through nuclear
matter is discussed. The close anology between the Landau - Pomeranchuk -
Migdal (LPM) effect in QED and the  emission of soft gluons, suggested
 in ref. \cite{BDPS} for ``hot" plasma, is  confirmed and the relation
between Mueller's approach and traditional calculations is  established.
 It is shown that  perturbative QCD can be applied to take into account the
LPM coherent suppression both for photon and gluon induced radiation.
The formulae for the photon and gluon radiation densities  are presented.

\end{titlepage}

\section{ Introduction }
The goal of this paper to study the Landau - Pomeranchuk - Migdal  (LPM) effect
 \cite{LPM} for the  emission of photons and gluons in  nuclear matter.
 We show that the typical distances  which characterize
the succesive rescaterring of quarks and gluons in  nuclear matter are
 small enough to justify the QCD approach to this case.
 We shall
 generalized the formalism suggested in refs. \cite{GW}\cite{BDPS} to the case
 of  nuclear matter using the QCD approach for the interaction
 of quarks and gluons.

We consider the emission of soft gluons ( photons ) with
  energies $ \omega \,\ll
\, E$, where $E$ is the energy of the projectile. The assumption that the mean
free  path $\la$ of the projectile is much larger than the screening radius
 in the nuclear matter, $ \la\,\gg\,\m^{-1}$, allows one to treat successive
 rescatterings as independent ( see ref. \cite{LPM} ) and simplifies
 all formulae reducing the problem to an eikonal picture of classical
 propagation of a relativistic particle with $E\,\gg\, \m$ through a medium.

As  is well known, the QED emission amplitude for a single scattering
 can be written in terms of a transverse velocity ${\vec{u}}_{\perp}\,=\,
\,\frac{{\vec{k}}_{\perp}}{\omega}$ where ${\vec{k}}_{\perp}$ is the transverse
 momentum of the photon with respect to the direction of the fast (massless)
 charged
 particle.  For the emission of a photon from the scattering center
$i$ it reads:
\beq
\vec{J}_{i}\,\,=\,\,\frac{{\vec{u_i}}_{\perp}}{u^2_{i \perp}}\,\,-\,\,
 \frac{\vec{u}_{i - 1 \perp}}{u^2_{i - 1, \perp}}\,\,;\,\,\,\,
\vec{u}_{i \perp}\,\,=\,\,\vec{u}_{(i - 1) \perp}\,\,-\,\,\frac{\vec{
q}_{i \perp}}{E}\,\,;
\eeq
 where
 $\vec{q_{i\perp}} $ is the momentum transfer to the scattering center $i$.

The LPM effect comes from the intenference between the amplitudes (1) in
the calculation of the inclusive cross section of photon production.
 The relative
 eikonal phase of the two amplitudes due to centers $i $ and $j$ in
 relativistic kinematics is equal to:
\beq
\p_{ij}\,\,=\,\,k^{\m}(x_i - x_j)_{\m}
\,\,=\,\,\sum^{j - 1}_{m = i}\,\,\frac{z_m - z_{m +
1}}{\tau_m(\o)}\,\,;\,\,\,\,
\tau_m\,=\,\frac{2\o}{(\vec{k}_{m \perp})^2}\,=\,
\frac{2}{\o u^2_{m \perp}}\,\,;
\eeq
where  $ \tau_m$ is the  radiation formation time
 and $z_m$ is the longitudinal coordinate of the
 $m$-th centre.

One can easily estimate the average phase \cite{LPM}, assuming that all
$z_m - z_{m + 1}$ are equal to the mean free path $\la$, and obtain:
\beq \label{PHI}
\p( n = j - i )\,\,\approx\,\,\frac{\la \o}{2} \,\sum^{j - i - 1}_{m = 0} \,\,
u^2_{i + m}\,\,\approx\,\,\frac{\la \o}{2}\,\,\{\, n u^2_{i \perp}\,\,+\,\,
\frac{n ( n - 1 )}{2} \,\frac{\m^2}{E^2}\,\}
\eeq
Here we have
 used the assumption that  the transverse momentum of the projectile,
  after $m$ collisions with a typical momentum transfer $\VEV{q^2_{\perp}}
\,=\,\m^2$, can be treated as a random walk in the transverse plane which
gives :
 $ u^2_{i + m,\perp}\,\,=\,\,u^2_{i \perp} \,\,+\,\,m \,\frac{\m^2}{E^2}$.
Since the amplitude of Eq.(1) is small at
$u^2_{i \perp}\,\,>\,\,\frac{\m^2}{E^2}$ the second term in \eq{PHI} dominates
 at large $ N\,\,\gg\,\,1$.
The cross section vanishes  if $\p \,>\,1$. It means that the number
 ( $n$ ) of scattering centres
that act coherently and  can be treated as a single radiator,
 is restricted from above by the value $N_{coh}( \o )$:
\beq \label{NCOH}
n\,\,\leq\,\,N_{coh}( \o )\,\,=\,\,2 \,\sqrt{\frac{ E^2}{\la \m^2 \o}}
\eeq
Considering a group of centeres with $ i - j \,<\,N_{coh}$ as a single
radiator, we can estimate the radiation density ( see refs. \cite{LPM}
\cite{BDPS} \cite{LPMREV} for details):
\beq \label{RADAPPR}
\o\,\frac{d^2 I}{ d \o  \,d z}\,\,=\,\,\frac{1}{\la} \,
\{\o \frac{d I}{ d \o}\}_{one\,\,centre}\,\cdot\,\frac{1}{N_{coh}( \o )}
\,\,\propto\,\,\frac{\alpha_{em}}{\la N_{coh}}
\eeq
One can see three  kinematic regions in which \eq{RADAPPR} gives different
answers for the radiation density:

1. $N_{coh}( \o )\,\,>\,\,1 $ or $\o \,<\,\o_{BH}\,=\,\frac{2 E^2}{\la
\m^2}\,=\,
E \cdot\frac{E}{E_{LPM}}$ where $ E_{LPM} \,=\,\frac{1}{2} \m^2 \la$.
Here
\beq
\o\,\frac{d^2 I}{ d \o  \,d z}\,\,\propto\,\,\alpha_{em} \,\sqrt{\frac{\m^2}
{\la E^2}\,\o}\,=\,\frac{\alpha_{em}}{\la} \,\sqrt{\frac{\o}{\o_{BH}}}
\eeq
This equation gives the Migdal result \cite{LPM}.

2. $N_{coh}(\o)\,\,\approx\,\,1$ or $\o \,\geq \o_{BH}$.
In this kinematic region each scattering centre radiates separately and
 we derive the well known Bethe - Heitler limit:
\beq \label{BH}
\o\,\frac{d^2 I}{ d \o  \,d z}\,\,=\,\,\frac{1}{\la} \,
\{\,\o \frac{d I}{ d \o}\}_{one\,\,centre}\,\propto \,\frac{\alpha_{em}}{\la}
\eeq

3.  In a medium with a final longitudinal size $L$, we have to assume
that $L$ is  large enough so as to have $N_{coh}$ successive interactions.
It means:
\beq \label{FINSIZE}
\la N_{coh}\,\,<\,\,L \,\,\,\,or\,\,\,\,
\o\,\,>\,\,\o_{fact}\,=\,\frac{\la E^2}{L^2\m^2}
\eeq
For $\o \,\
<\,\o_{fact}$  one recovers the factorization limit in which the whole
medium radiates as one centre with an amplitude
\beq
\sum^n_{i=1}\,\vec{J_i}\,\,=\,\,\frac{\vec{u}_{n\perp}}{u^2_{n \perp}}\,\,
-\,\,
 \frac{\vec{u}_{1 \perp}}{u^2_{1 \perp}}\,\,;
\eeq
At first sight, the factorization region is irrelevant for the emission in a
nuclear matter. However, it is not so, since each produced quark or gluon
lives only a finite time ( $ \tau_{max} \,\propto\,
\frac{E}{\VEV{p^2_{\perp}}}$
 where $p_{\perp}$ is the transverse momentum of the
produced charged particle which
 embodies successive rescattering)
 inside a nuclear matter. Substituting $ L = \tau_{max}$ in \eq{FINSIZE}
we obtain $\o_{fact}\,=\,\frac{\la \VEV{p^2_{\perp}}}{\m^2}$.
Therefore we expect the LPM effect for the photon energy range
\beq
\frac{\la \VEV{p^2_{\perp}}}{\m^2}\,=\,\o_{fact}\,<\,\o\,<\,\o_{BH}\,
=\,\frac{E^2}{E_{LPM}}
\eeq
The situation becomes even more interesting in the case of gluon emission.
As was discussed in ref.\cite{BDPS}, the produced gluon itself embodies the
 successive rescatterings in a medium. The life time of the
 gluon is $\tau_{max}\,
\propto\,\frac{\o}{k^2_{\perp}}\,\,\approx\,\,\frac{1}{\o}$.  To discuss
the contribution of  gluon rescattering to the LPM effect we have to demand
 that
\beq
\la N_{coh} ( \o ) \,\,<\,\,\tau_{max} ( gluon ) \,\,\approx \,\,\frac{1}{\o}\,
\,\,\,or\,\,\,\,
\o \,\,<\,\,\frac{1}{\la^2 \o_{BH}}
\eeq
Therefore~~ we~~ anticipate~~ for~ gluons~~ the classic~~ LPM
 suppression~~ only~~ for
$\o \,\,<\,\,min\{\,\o_{BH},\frac{1}{\la^2 \o_{BH}}\,\}$
and\,~~ some ~~\,\,new behaviour of\,\,~~ the radiation \,\, density\,\,~~ for
$
min\{\,\o_{BH},\frac{1}{\la^2 \o_{BH}}\,\}$
$ \,\,\,<\,\,\,\o\,\,<\,\,\,$
$ max\{\,\o_{BH},\frac{1}{\la^2 \o_{BH}}\,\}
$.

{}From the above sketch of the physics of the LPM effect it is clear that
there are a number of questions that we have to answer: (i)
what is the value of
 the mean free path $\la$ in QCD; (ii)
 what is the correct procedure of averaging
over transverse momentum and what is the value of $\m$; and, (iii) what kind of
 LPM supression do we expect for gluon emission. We will answer these
questions in this paper, starting with the photon emission
 from  a fast quark in a
nuclear matter in the second section.
In the third section we shall consider the interconnection
 between the LPM effect
 and the Mueller technique \cite{MUT}, which was widely used to calculate
 the inclusive spectra of produced particles.
In the fourth section we  discuss  the LPM like supression for gluon emission
 in a nuclear matter.
A summary and final discussion are presented in the  conclusion.

In general, we attempt to adjust our notation and normalization to agree with
 those  of Baier et al \cite{BDPS}, using $E$ and $p$ for
 the energy and momentum of a projectile which propagates through
a nuclear matter;
$\o$ and $k$ for the energy and momentum of a radiated photon (gluon) and
$q_{i \perp}$ for the  momentum transfer to the  $i$-th scattering centre.

\section{ The LPM effect for photon radiation}
The inclusive soft photon spectrum from
a fast quark with energy $E$ can be written in the form ( see for example
 \cite{BDPS}):
\newpage
\beq  \label{INCSP}
 \omega\frac{dI}{d\omega\,d^2 u} =
\frac{\alpha_{em}}{\pi^2}
\VEV{ \sum^{N}_{i=1}\,\vert \vec{J}_i e^{i\,k_{\mu} x^{\m}}\vert^2}\,\,=
\eeq
$$\frac{\alpha_{em}}{\pi^2}
\VEV{ 2 \real \,\sum^{N}_{i=1} \sum^{N}_{j=i+1} \, \vec{J}_i \vec{J}_j
\left[\,  e^{ik_{\mu} (x_i-x_j)^{\mu}}\,\, -1\,\right]\,\,+\,\,
 \vert \sum^N_{i=1} \vec{J}_i\vert^2  }
$$
The brackets $\VEV{\,\, }$ indicate the averaging procedure discussed in
\cite{GW} \cite{BDPS}, and we shall discuss it below in detail.
The differential energy distribution of photons radiated
{\em per unit length}\/ is given by \cite{BDPS}
\beq \label {ENSP}
 \omega\frac{dI}{d\omega\, dz} =
\, \frac{\alpha_{em}}{\pi}\,
\int\frac{d^2 U_0}{\pi} \VEV{ 2\real\, \sum^{\infty}_{n=0}
 \vec{J}_1 \vec{J}_{n+2}
\left[\,
\exp\left\{ i \frac{1}{\tau} \sum^{n + 1}_{l=1}\ U^2_l\,(
\,z_{l +1}-z_l\,)\,\right\}\,\,\, -1 \,\right] }.
\eeq
Here we have expressed the relative phases in terms of the
 rescaled  transverse velocity,
\beq\label{UDEF}
 \vec{U}_{l} \equiv \vec{u}_{l} \cdot E\,\,=\,\,\frac{\vec{k}_{\perp}}{x_F}
\,\,-\,\,\sum^{l}_{i = 1} \,\vec{q}_{i\,\perp}\,\,;\,\,\,\,\,
 \vec{U}_{0}\,\,=\,\,\frac{\vec{k}_{\perp}}{x_F}
\eeq
where  $x_F \,\,=\,\,\frac{\o}{E}$ is the fraction of quark energy
carried by the photon.
We  also introduced the characteristic parameter
\beq \label{TAUDEF}
\frac{1}{\tau}\,\, = \,\,\frac{\o}{2\,E^2}
\eeq
Now we shall
 define the averaging procedure that has to be implemented  in \eq{ENSP},
starting with the simplest problem, namely, with the propagation of the
 quark through a nuclear matter. This problem was solved a long time ago
\cite{LR1}.
Indeed, let us consider the inclusive spectrum of a fast quark
produced at a point $z$, with respect to
 its transverse momentum ( $p_{\perp}$ )
 after $n$ rescatterings with nucleons at points
 $z_1,z_2,
...z_n$. This spectrum can be written as \cite{LR1} ( see Fig.1)
$$
\frac{d \s_n}{d^2 p_{\perp}}\,\,=\,\,
e^{- \s_{tot}\,(\, z_1 - z\,)} \,d z_1 \,\rho\,\frac{d \s}{ d^2 q^2_{1 \perp}}
\,\cdot\,e^{- \s_{tot}\,(\, z_2 - z_1\,)} \,d z_1
\,\rho\,\frac{d \s}{ d^2 q^2_{2 \perp}}\,...\,
e^{- \s_{tot}\,(\,z_1 + L - z_n\,)} \,d z_n \,\rho\,\frac{d \s}
{ d^2 q^2_{n \perp}}
$$
\beq \label{QUARKSP}
\,\,\delta^{( 2 )} \(\,\sum^{n}_{m = 1} \,{\bf \vec{q}_{m \perp}\,\,-\,\,
\vec{p}_{\perp}\,}\) \,\prod_{m} \,d^2 q_{m \perp}\,\,\,\,\,\,(\,z_1\,<\,z_2
\,<...<\,z_n \,\,<\,z_1 + ( L\,=\,\tau_{max})\,)
\eeq
Here, $\rho$ is the nucleon density, $\frac{d \s}{ d^2 q_{m \perp}}$ is the
spectrum of the quark due to rescattering with momentum transfer $q_{m \perp}$
 which  can be written through the unintegrated  parton density $\phi$
 \cite{LR1} ( see also relevant papers \cite{LR2} \cite{MU90}):
\beq \label{QCRSEC}
\frac{d \s}{ d^2 q_{m \perp}}\,\,=\,\,\frac{8\pi^2 \as^2 ( q^2_{ m \perp} )}{ 9
\,
q^4_{m \perp}}\,\cdot\,( x \phi_N (x, q^2_{n \perp}))
\eeq
where $ x \phi_N( x, q^2_{\perp} )$ is the nucleon parton density
  with \,\,\,$x =
 \frac{1}{\tau_{max}}$. The relation between  the unintegrated parton
 density  $\phi$ and the nucleon's gluon structure function
 $x G_N (x, q^2_{\perp})$ can be calculated using the following
 equation:
$$
\as(Q^2)\,\cdot\,x G(x, Q^2)\,\,=\,\,\int^{Q^2}\,\,d\,q^2_{\perp}\,
\as(q^2_{\perp})\,\,\phi (x, q^2_{\perp})
$$
 The factor $ exp ( - \s_{tot}$\,\,
$ \,(\, z_m\, -\,z_{m - 1}\,)\,)$ in \eq{QUARKSP}
describes the fact that the quark has no  inelastic
collisions between  the centres $z_{m - 1}$
 and $ z_m$. The total cross section is equal to $\s_{tot} \,=\,\int
\,d^2 q_{m \perp} \,\frac{d \s}{ d^2 q_{m \perp}}$ and it is formally
 divergent at $q_{m \perp} \,\ra\,0$. However, this divergency cancels  against
the divergency of the inelastic cross section. To see this fact, it is
convenient to describe the process in the transverse coordinate space
 $r_{\perp}$ using the well know representation of the $\delta$ function:
$$
\int \,\frac{d^2 r_{\perp}}{ ( 2 \pi )^2} \,\exp[ i \, \sum_m
{\bf\( \vec{q}_{m \perp} \,\,-\,\,\vec{p}_{\perp}\,\)\,\cdot
\,\vec{r}_{\perp}}]
\,\,=\,\,\delta^{( 2 )} \(\,\sum^{n}_{m = 1} \,{\bf \vec{q}_{m \perp}\,\,-\,\,
\vec{p}_{\perp}\, }\)
$$
Summing over $n$ we obtain
\beq \label{INCQUARK}
\frac{d \s}{d^2 p_{\perp}}\,\,=\,\,\int \,\frac{d  r^2_{\perp}}{  4\, \pi^2  }
\,\,J_0 (\, p_{\perp}\,r_{\perp}\,)\,\,\exp\( -\,\s(\, r^2_{\perp}\, )\,
\rho\,L\,\)
\eeq
with
\beq  \label{QUARKCRSEC}
\s ( r^2_{\perp})\,=\,\int d q^2_{\perp}
\,\{\, 1\, - \,J_0 (\, q_{\perp}\,
r_{\perp}\, \}\,\frac{d \s}{ d q^2_{\perp}}
\,\,=\,\,\frac{\as(\frac{4}{r^2_{\perp}})}{3}
\,\,\pi^2\,\,r^2_{\perp}\,\,
\(\,\, x G_N( \frac{4}{r^2_{\perp}},x )\,\,\)
\eeq
Let us investigate $ \frac{d \s}{d^2 p_{\perp}}$, adopting the following method
 of integration over $r_{\perp}$ in \eq{INCQUARK}:(i)  we consider
 $xG_N(\frac{4}{r^2_{\perp}}, x)$ as a smooth function of $r^2_{\perp}$ since
it depends on $r^2_{\perp}$ only logarithmically; (ii) we integrate over
$r_{\perp}$  taking only into account  $\s \,\propto \,r^2_{\perp}$; and,
(iii) we put in $xG_N(\frac{4}{r^2_{\perp}}, x)$
  $r^2_{\perp} \,=\,r^2_{0 \perp}$, where $r^2_{0 \perp}$ is the typical value
of $r^2_{\perp}$ in the integral.
The result of the  integration is obvious, namely
\beq \label{RDEP}
\frac{d \s}{d^2 p_{\perp}}\,\,=\,\,\frac{1}{ 4 \,\rho\,L\,\,
\pi^2\,[\, \frac{\as \pi^2}{3}
x \,G_N ( \frac{4}{r^2_{0 \perp}}, x)\,]}\,\,e^{ -\,\,
\frac{p^2_{\perp}}{4 \,\rho\,L\,\, \frac{\as \pi^2}{3}
x \,G_N ( \frac{4}{r^2_{0 \perp}}, x)}}
\eeq
 One can find the value of $r^2_{0 \perp}$  by calculating the saddle point in
$r_{\perp}$ integration, which gives:
\beq \label{TYPR}
r^2_{0 \perp}\,\,=\,\,\frac{p^2_{\perp}}{ 4 \,( \,\rho\,L\,\,
 \frac{\as \pi^2}{3}
x \,G_N ( \frac{4}{r^2_{0 \perp}}, x)\,)^2} \,\,\,\,for\,\,\,\,
p^2_{\perp}\,r^2_{0 \perp}\,\,\gg\,\,1\,\,;
\eeq
$$
r^2_{0 \perp}\,\,=\,\,\frac{1}{ 2 \,( \,\rho\,L\,\,
 \frac{\as \pi^2}{3}
x \,G_N ( \frac{4}{r^2_{0 \perp}}, x)\,)} \,\,\,\,for \,\,\,\,
p^2_{\perp}\,r^2_{0 \perp}\,\,\ll\,\,1\,\,;
$$
We can trust our calculation in pQCD only if $r_{\perp} \,\,\ll\,\,r_{soft}$,
where $r_{soft} $ is the  scale at which the nonperturbative QCD corrections
 become essential. It means that the value of $p_{\perp}$ should be
smaller than $ 2 \,( \,\rho\,L\,\,
 \frac{\as \pi^2}{3}
x \,G_N ( \frac{4}{r^2_{0 \perp}}, x)\,)$. For bigger $p_{\perp}$, in
 \eq{INCQUARK}, we take $r_{\perp} \,\propto \,\frac{1}{p_{\perp}}$ which gives
$\frac{d \s}{d^2 p_{\perp}}\,\,\propto\,\,\frac{1}{p^2_{\perp}}\,
\s(\frac{1}{p^2_{\perp}})\,\, \rho\,L$.
The main message from this calculation is that
the typical distances that work in the rescattering inside the nuclear matter
 turns out to be small. This justifies the use of pQCD  in
this case. Therefore, we can make a guess that the parameter $\m$ that has been
discussed in the introduction is equal to $\m^2 \,=\,\frac{1}{r^2_{0 \perp}}$.
However, we have to be very carefull because the $p_{\perp}$ - distribution
 has a tail at large values of $p_{\perp}$. We will show that we can safely use
\eq{RDEP} to study the LPM effect.

The value of the mean free path $\la$ can  also be estimated from
\eq{INCQUARK} and it is equal $\la \,=\,\frac{1}{\s(r^2_{0 \perp}) \,\rho}$.
This is clear if we rewrite \eq{INCQUARK} as
\beq \label{AVER}
\frac{d \s ( \D z)}{ d^2 p_{\perp} \,d z} \,\,=\,\,
\int \,\frac{d  r^2_{\perp}}{  2  }\,\s( r^2_{\perp})\,\rho\,
J_0 (\, p_{\perp}\,r_{\perp}\,)\,\exp\( -\,\s(\, r^2_{\perp}\, )\,
\rho\,\D z\, \)
\eeq
where\,\,\, $ \D z$\,\,
{}~~ is ~~the longitudinal~~ distance ~~that ~~ the quark~~  passes ~in ~a ~
nuclear
{}~~ matter (\,$ \s(\, r^2_{\perp}\, )\,
\rho\,\D z\,$~~~$ \,\,>\,\,1$\,).

 Actually, \eq{AVER} sets the averaging procedure for \eq{ENSP}.
Indeed
\beq \label{AVER1}
\VEV{\,\,}\,\,=\,\,
\prod^{n + 2}_{ l = 1} \,\frac{d^2 U_{l \perp}}{\pi}\,\prod^{n+2}_{l = 1}
 \, d z_l\,\,\frac{d \s (z_l -  z_1)}{ d^2 p_{l\,\perp} \,d z}
\,\,;\,\,\,
 z_1\,<\,z_2\,...\,<\,z_{l - 1}\,<\,z_{l}\,<\,...<\,z_{n+2}\,
<\,z_1 \,+\,L
\eeq
Using the explicit expression for $\vec{J}_{i}$ of eq.(1), one can see, after
integration over the azimuthal angle,
 that only big transfer momenta $q_{i \perp}
\,\,\gg\,\,\frac{k_{\perp}}{x_F}$ contribute to \eq{ENSP}. It means that
$ p_{l \perp} \,\,\ra\,\,U_l$. This fact allows us to make the integration over
$U_l$ in \eq{AVER1}  explicitly which leads to
\beq \label{UINT}
\VEV{ \vert \frac{d I_l}{d z_l}\vert}\,\,\equiv\,\,
\int^{\infty}_{U^2_0}\,\,d\, U^2_{l \perp}\,\,
\frac{d \s (z_l -  z_1)}{ d^2 U_{l}\,\,d \,z_l}
\,\,\exp\left\{ i \frac{1}{\tau}  U^2_l\,(
\,z_{l +1}-z_l\,)\,\right\}
\eeq
Introducing a new variable $\ti z_l \,\,=\,\, \,\rho\,z_l\,\,{\bf D}\,\,=\,\,
\rho\,\,z_l\,\,
 \frac{\as \pi^2}{3}
x \,G_N ( \frac{4}{r^2_{0 \perp}}, x)$ one can take the integral over $U_l$
 and obtain:
\beq \label{AVERU}
\VEV{ \vert \frac{d I_l}{d z_l}\vert}\,\,=\,\,{\bf D}\,\rho\,\frac{d}{d \ti
z_l}
\,\{\,\frac{1}{i \,\kappa \,\ti z_l\,\Delta \ti z_l\,-\,\frac{1}{4}}\,\cdot\,
\exp[\,-\,\frac{U^2_0}{4 \ti z_l}\,\,+\,\,i\,\kappa\,U^2_0\,\Delta \ti z_l\,]
\,\}\,=
\eeq
$$
\,=\,\,\{\,16\,i\,\kappa\,\Delta \,\ti z_l\,\,-\,\,\frac{U^2_0}{{\ti
z_l}^2}\,\}
\,\cdot\,\exp[\,-\,\frac{U^2_0}{4 \ti z_l}\,\,+\,\,
i\,\kappa\,U^2_0\,\Delta \ti z_l\,]
$$
where $\kappa \,=\,\frac{1}{\tau\,\cdot\,{\bf D}\,\rho}$ and $\Delta \ti z_l
\,=\, \ti z_{l + 1}\,-\,\ti z_l$. We anticipate that $\kappa \ti z_l
 \Delta \ti z_l\,\,\ll\,\,1$ and have expanded the answer with respect to this
 parameter.

The above equation allows us to obtain the functional equation for
\beq \label{DEFP}
\Phi_n (U^2_0, z )\,\,=\,\,\prod^{n}\,\,\VEV{ \vert \frac{d I_l}{d z_l}\vert}
\eeq
Namely,
\beq \label{FUNCEQ}
\Phi_{n - 1} (U^2_0, z - \Delta z)\,\{ \,16 i \kappa \Delta z \,\,-\,\,
\frac{U^2_0}{ z^2_l}\,\}\,\cdot\,\exp\,
[\,-\,\frac{U^2_0}{4 z}\,\,+\,\,i \,\kappa
\,\Delta z\,]\,\,=\,\,\Phi_{n} (U^2_0, z)
\eeq
Considering $\Delta z \,\ll\,z$, we can solve \eq{FUNCEQ} and obtain
\beq \label{SOLPHI}
\Phi_{n}(\kappa,U^2_0, z)\,\,=\,\,
\left( -\frac{U^2_0}{ {\ti z}^2}\right)^{n + 1}
\,\exp\,[\,-\,\frac{U^2_0}{4 \ti z}\,\,(\,n\,+\,1\,)
\,\,+\,\,i\,\kappa U^2_0\,\ti z\,\,+
\,\,\frac{16 i \kappa\,{\ti z}^3}{3 \,U^2_0}\,]
\eeq
Substituting \eq{SOLPHI} in \eq{ENSP}, summing over $n$ and substructing  the
value of the integral at $\kappa = 0$, which corresponds  to 1 in \eq{ENSP},
we obtain the answer:
\beq \label{ANSWER1}
\o\,\frac{d I}{ d \o d z}\,\,=\,\,
\frac{\alpha_{em}}{\pi}\,4\,{\bf D}\,\rho\,
\int \,\frac{d U^2_0}{U^2_0}\,\int d \ti z
\,\,\frac{sin^2\{\frac{\kappa}{2}\,[\,\frac{16 {\ti z}^3}{3\,U^2_0}\,\,+\,\,
U^2_0\,\ti z\,]\,\}\,( \,- \frac{U^2_0}{{\ti z}^2}\,)^2\,
e^{-\,\frac{U^2_0}{2 \ti z}}}
{1\,\,+\,\,\frac{U^2_0}{{\ti z}^2}\,e^{-\,\frac{U^2_0}{4 \ti z}}}
\eeq
Introducing  a new variable $\xi \,=\,\frac{U^2_0}{\ti z}$ one can rewrite
\eq{ANSWER1} in the form:
\beq \label{ANSWER2}
\o\,\frac{d I}{ d \o d z}\,\,=\,\,
\frac{\alpha_{em}}{\pi}\,4\,{\bf D}\,\rho\,
\int \,\,\frac{d U^2_0}{U^2_0} \,\,\int^{\infty}_{\xi_0} \,\,\xi^2\,\,d \,\xi
\,\,\frac{sin^2(\frac{\kappa\,U^2_0}{2}\,[\frac{16}{3 \xi^3}\,+\,\frac{1}{\xi}]
\,e^{\,-\,\frac{\xi}{2}}}{U^2_0\,\,+\,\,\xi^2\,e^{\,-\,\frac{\xi}{4}}}
\eeq
The lower limit in  the
$\xi$ integral is  $\xi_0\,\,=\,\,\frac{U^2_0}{\rho\,
{\bf D}\,L}$. For $\kappa \,U^4_0\,\,\gg\,\,1$   $\xi \,\approx\,1$ contributes
 to the integral and we recover the BH limit. For $\kappa\,U^4_0\,\,\ll\,\,1$
 the main contribution comes from the region of small $\xi$, but $\xi\, >\,
\xi_{min}\,=\,( \frac{8 \kappa U^4_0}{3} )^{\frac{1}{3}}$. In this  region
  the argument of the  $sin$ in \eq{ANSWER2} turns out to be small. Expanding
$ sin$ and  doing all integrations we get
\beq \label{FINANS}
\o\,\frac{d I}{ d \o d z}\,\,=\,\,
\frac{\alpha_{em}}{\pi}\,4\,{\bf D}\,\rho\,\frac{8}{9}\,\,\sqrt{\kappa}\,\,
=\,\,\frac{32\alpha_{em}}{9\,\pi}\,\,\sqrt{\frac{\o \,{\bf D}\,\rho}{2\,\,E^2}}
\eeq
To obtain the final answer we need to specify the argument of the gluon
structure function in ${\bf D}$. It turns out that the value of the typical
$r^2_{0 \perp}\,\,\propto\,\sqrt{\kappa}$ and it is small  for small $\kappa$
for which we have
 derived \eq{FINANS}. Therefore we are  justified  in  our approach and the
 final answer is
\beq \label{FINANS1}
\o\,\frac{d I}{ d \o d z}\,\,=\,\,
\frac{32\alpha_{em}}{9\,\pi}\,\,\sqrt{\frac{\o \,\frac{\as \pi^2}{3}
\,\,( x\,G(x, r^2_{soft} \frac{1}{\sqrt{\kappa}})\,\rho}{2\,\,E^2}}
\eeq
The region of applicability of \eq{FINANS1} can be found from the condition
that the typical distances in our calculations are small so as
 to use perturbative
calculation. In other words, it can be found
 from  the integral of \eq{ANSWER2} in the region
$\xi\,\approx \,1, \kappa\,U^4_0 \,\gg\,1$. It gives
\beq \label{BHQCD}
\o_{BH}\,\,=\,\,\s(r^2_{soft})\,\rho\,E^2
\eeq
where $r^2_{soft}$ is the scale from where we can use perturbative QCD
 ($r^2_{\perp} \,<\,r^2_{soft}$), and which
 should be calculated in a nonperturbative QCD approach.

The factorization limit comes from the contstraint that $\xi_{min}
\,\leq\,\xi_0$
and gives
\beq \label{OFACT}
\o_{fact}\,=\,\frac{\VEV{p^2_{\perp}}}{\rho \s(\frac{1}{\VEV{p^2_{\perp}}})}
\eeq
Summarizing this section we want to mention that we did not obtain the
$\\sqrt{kappa}\,
 ln \kappa$ behaviour of the radiation density in a nuclear matter
 as it was done for ``hot"
deconfined plasma in ref.\cite{BDPS}. The difference comes from the averaging
 over momentum transfers $\vec{q}_{l \perp}$ which turns out to be quite
 different from the screened Coulomb potential applied in ref.\cite{BDPS} for
a ``hot" plasma.
\section{The LPM effect and Mueller technique.}
In this section we are going to discuss the interrelation between our approach
and the Mueller's technique \cite{MUT},
which, together with the AGK cutting rules \cite{AGK},
remarkably simplify the calculation of the inclusive spectra of produced
 particles. It is  widely used in all
Reggeon inspired calculations. The Mueller diagrams for
the inclusive spectra of emitted photons are shown in Fig.2.
There are two main assumptions that make the Mueller technique simple and
 attractive: (i) the spectrum of a produced particle does not depend on
 the kinematics of the incoming one; and, (ii) the longitudital part of the
momentum transfer turns out to be so small at high energy that it
 can be neglected.

The first assumption holds for the emission of sufficiently soft particles
and it corresponds to the factorization properties of the production of
particles from sufficiently small distances in QCD \cite{FACTOR}. At first
sight
one can prove that the diagrams of Fig.2b do not contribute to the inclusive
 cross section using the AGK cutting rules and factorization. However, we are
dealing with the emission from the fastest particle and factorization does not
work in this case. As a result the sum of the diagrams of Fig.2 gives the
factorization limit, or in other words the second term in \eq{INCSP}. We plan
 to discuss this term in a separate paper where we shall consider the
 inclusive cross section for the emitted photon or gluon. However, even
 the sum of all Mueller diagrams cannot reproduce the LPM effect.

The LPM effect comes from a more carefull consideration of the dependance of
the
 production amplitude on the longitudinal part of  the momentum transfer
  $q_l$. From Fig.3 we can see that the
 interaction with the  centre $`l'$ in the first term of \eq{INCSP} has the
 longitudinal component of the momentum transfer $q_z$ ($z$ is the direction
of the incoming quark) which is equal:
\beq \label{LONGQ}
q_z\,\,=\,\,\frac{1}{ 2  E}\,\cdot\,\{\,(\,p\,+\,k\,)^2\,\,-
\,\,(\,p'\,+\,k\,)^2\,\}\,\,=\,\,\frac{1}{2 E}\,\cdot\,\{\,2\,k_{\mu}
 \,q_{l \mu}\,\}\,\,=\,\,\frac{\o}{2 E^2}\,\{\,U^2_{l + 1} \,-\,U^2_l\,\}
\eeq
where we have
neglected the change in the energy of the fast quark due to
the photon emission, as well as due to rescattering.
 On the other hand, the longitudinal
part of the momentum $q'_l$ is small.  At high energy we are doing all
 calculations for quark - nucleon interactions in the leading log(1/x)
approximation ( see for example  ref.\cite{GLR} ), in which we neglect
 the longitudinal momentum dependance for all gluon and quark propagators.
The only dependance on the longitudinal momentum comes from
 the fact that the total
 momentum transfer for the scattering off the
 $l$-th centre ($Q_l \,= q_l + q'_l$)
differs from zero and it is equal $Q_{z\,l}\,\,=\,\,q_{z\,l} $. Considering
 nucleons, in a nuclear matter, as nonrelativistic particles, we have neglected
 the change of the energy component of $Q_l$. Finally, we have for each
 rescattering
\beq \label{PSIFUNC}
\Psi^{*}_{initial}( l ) \,\cdot\,\Psi_{final} (l )
\,\,=\,\,\rho\,\,e^{i\,Q_{zl}\,z_l}\,\,=\,\,\rho\,e^{i \frac{1}{\tau}\,
(\, U^2_{l + 1}\,-\,U^2_l\,)}\,\,,
\eeq
where $\Psi(l)$ is the wave function of the $l$-th nucleon in a nuclear matter.
One recognizes the phase that we have taken into account in \eq{ENSP}.

Therefore, the LPM effect can be derived from Mueller diagrams if one takes
 into account the dependance on the longitudinal part of the momentum transfer.
We will show that the Mueller diagrams
will be very useful to reach simple and transparent understanding of the
 LPM - like effect in the case of  gluon emission.
\section{The LPM effect for gluon emission.}
In this section we consider the gluon radiation in a nuclear matter using the
Mueller diagrams of Fig.3. To write down the diagrams of Fig.3 we have to
 specify the expression for $\vec{J}_{i}$ for gluon emission and calculate
 the longitudinal part of the momentum transfer (see \eq{LONGQ} ).

The first
 ingredient in the soft limit (\, $\o\,\,\ll\,\,E$ \,)
 was calculated many years ago by Lipatov and collaborators
\cite{BFKL} and has been confirmed  using quite different technique
(see ref. \cite{GW} in which such a calculation was done just for the case of
induced gluon radiation in a medium.  It is very relelevant to our
approach). The answer is (see Fig.4):
\beq \label{JQCDGLUON}
\vec{J}_{i}\,\,=\,\,\frac{\vec{k}_{\perp}}{k^2_{\perp}}\,\,-\,\,
\frac{ \vec{k}_{\perp}\,\,-\,\,\vec{q}_{l \,\perp}}{ (\,\vec{k}_{\perp}\,\,-
\,\,\vec{q}_{l \,\perp}\,)^2}
\eeq
with  the colour factor which is equal to  the colour factor of the Feynman
 diagram of Fig.4(2). To simplify  the calculation of the colour factor we
 perform them for QCD with a large
 number of colours $N_c$ neglecting all terms
of the order of $1/N_c$.

For a gluon we have two rescatterings which are shown in Fig.4b and Fig.4c.
The value of $q_{l z}$, for the quark rescattering,  has been calculated (
 see \eq{LONGQ} ) and it is small in the soft region where $E\,\gg\,\o$.
For the gluon rescattering $q_{lz}$ is equal to:
\beq \label{LONGG}
q_z\,\,=\,\,\frac{1}{ 2  E}\,\cdot\,\{\,(\,p\,+\,k\,)^2\,\,-
\,\,(\,p\,+\,k'\,)^2\,\}\,\,=\,\,\frac{1}{2 E}\,\cdot\,\{\,2\,p_{\mu}
 \,q_{l \mu}\,\}\,\,=\,\,\frac{\o}{2 \o}\,\{\,U^2_{G,l + 1} \,-\,U^2_{G,l}\,\}
\eeq
where
\beq \label{UGDEF}
 \vec{U}_{G,l}\,\,=\,\,\vec{k}_{\perp}
\,\,-\,\,\sum^{l}_{i = 1} \,\vec{q}_{i\,\perp}\,\,;\,\,\,\,\,\vec{U}_{G,0}\,
\,=\,\,\vec{k}_{\perp}\,\,.
\eeq
Comparing \eq{LONGQ} and \eq{LONGG} one can see that $q_{lx}$ due to quark
 rescattering is much smaller than $q_{l z}$ for a gluon collision. Therefore,
we can neglect the quark rescattering in the first approximation to the
 problem.

Therefore, we can write an equivalent
 expression to the one in \eq{ENSP} for gluon radiation
density, using the following substitutes:
\beq \label{SUBS}
\vec{U}_{l}\,\,\,\ra\,\,\vec{U}_{G,l}\,\,\,;\,\,\,\kappa\,\,\,\ra\,\,\,
\kappa_G\,\,\,=\,\,\,\frac{1}{2 \,\o\,{\bf D}\,\rho}\,\,.
\eeq
We have to make two comments:

1 .  Inspite of the fact that we are dealing with
 gluon rescattering in the averanging procedure over the transverse
momentum of the  produced gluon  $k_{\perp}$ given by \eq{AVER1}, we
 should use the same quark cross section given by
 \eq{QUARKCRSEC}. Indeed, the gluon -
quark pair scatters in the medium and the transverse separation between them
 due to many rescatterings off nucleons in the limit
 $q_{lz}\,\,\ra\,\,0$, as we
will show below, is of the order $r^2_{\perp}\,\,\propto \,\,\sqrt{\kappa_G}$.
The average momentum transfer in a single collision $\VEV{q_{l \perp}}$ is
of the order of $\VEV{q^2_{l \perp}}\,\,\approx\,\,
\frac{1}{r^2_{\perp}\,N_{col}}\,\,\ll\,\,\frac{1}{r^2_{\perp}}$, where
$N_{col}$
is the number of collisions which is big enough
 $\propto \,\,\frac{1}{\sqrt{\kappa_G}}$. Therefore, each gluon in the nucleon
cross section carries a transverse momentum $q_{l \perp}$ which is much
smaller than $\frac{1}{ r_{\perp}}$. It means that such a gluon interacts with
the total colour charge of the  gluon - quark pair, which is equal to the
charge
 of the quark at $N_c\,\gg\,1$.  The typical time that a gluon lives is
$\tau_G \,=\,\frac{\o}{U^2_0}$ which is much smaller than the life-time of
the quark $\tau_Q\,=\,\frac{E}{\VEV{p^2_{\perp}}}$. It means that for each
emitted gluon the quark plays the role of a  spectator that neutralizes
half of the gluon colour charge.

2. It was shown in ref.\cite{GW}
( see also ref. \cite{BDPS})  that quark rescattering contributes to the
 radiation density of gluons, if one takes into account  the
 $1/N_c$ corrections.
The origin of this contribution is the dynamic gluon correlations that have
 been studied in ref. \cite{HT}. Such correlations change the Glauber-type
 formula  of \eq{INCQUARK} that was used for  the
averaging in the nuclear matter.

The final answer is \eq{ANSWER2} with the substitutions
defined in  \eq{SUBS}. The value of
$\xi_0$ is equal to $\xi_0 \,=\,\frac{U^2_0}{\rho \,{\bf D} \,L}\,=\,
2 \kappa_G\,U^2_0$. For $\kappa_G U^4_0\,\,\ll\,\,1$,  we  derive the answer:
\beq \label{ANSWERG}
\o\,\frac{d I_G}{ d \o d z}\,\,=\,\,
\frac{ N_C \alpha_{S}}{2 \pi}\,4\,{\bf D}\,\rho\,\frac{8}{9}\,\,\sqrt{\kappa}
\,\,
=\,\,\frac{16 N_c \alpha_{S}}{9\,\pi}\,\,\sqrt{\frac{ \,\frac{\as \pi^2}{3}
\,\,( x\,G(x,  \frac{1}{r^2_{soft}\,\,\sqrt{\kappa}})\,\rho}{2\,\,\o}}
\eeq
$r^2_{soft}$ in \eq{ANSWERG} is a scale  of the ``soft" interaction.
 We can trust a
 perturbative calculation for the nucleon gluon distribution only for
$r_{\perp}
\,<\,r_{soft}$.
 The value of $1/x$ in \eq{ANSWERG} is equal to
$\frac{\tau_G}{\tau_{soft}}$ where $\tau_G\,=\,\frac{\o}{U^2_0}\,\,=\,\,
\sqrt{\kappa_G} \,\o$ and $\tau_{soft}$ is a typical time for the ``soft"
processes which we cannot specify in the leading log (1/x) approximation
(LL(1/x)A)
 which we have used to obtain the answer. To trust the
 LL(1/x)A we have to assume
that  thr  emitted gluon  energy
is so big that $\as ln \frac{\tau_{soft}}{\tau_G}
\,\,\geq\,\,1 $. It means that $\sqrt{\kappa_G}\,\o \,\,\gg\,\,\tau_{soft}$.

The  value of $\o_{BH}$ can be obtained from  the equation $\sqrt{\kappa }
\,\,=\,\,
 r^2_{soft} $  which gives
\beq \label{OBHG}
\o^{G}_{BH}\,\,= \,\,\frac{1}{r^2_{soft}\,\,\s (r^2_{soft})\, \,\rho}
\eeq
It should be stressed that unlike the QED case we cannot find the
 factorization limit for induced gluon emission. The physical reason for this
is
obvious since the phase of the propagating gluon cannot be small due to
its rescattering in a nuclear matter, at least in the kinematic region where
$\kappa_G$ is small. We would like to recall that we cannot trust our formulae
 for big valus of $\kappa_G$ as has been discussed above.

The radiation density of \eq{ANSWERG} is very close to the result of ref.
\cite{BDPS} and quite different from other attemps to estimate the LPM effect
for gluon radiation \cite{GW} \cite{RY}. The main difference between
\eq{ANSWERG} and the result of ref. \cite{BDPS} is the fact that the gluon
 nucleon density
 enters our  answer while the radiation density of ref. \cite{BDPS}
is proportional to $\frac{1}{\sqrt{\o}} \,\ln \o$. Since HERA data \cite{HERA}
shows sufficiently steep energy behaviour of the gluon structure function
\cite{HERA} which could be parametrized as $\frac{1}{x^{\o_0}}$ with the
value of $\o_0 \,\,\sim \,0.3 - 0.4$, the $\o$ behaviour of
the  gluon  radiation density can be evaluated as
 $\o\,\frac{d I_G}{ d \o d z}\,\,\propto\,\,\o^{-\,
\frac{1}{2}\,+\,\frac{\o_0}{4}}
$. The energy losses of the fast quark due to gluon emission  can be estimated
 integrating \eq{ANSWERG} over $\o$ up to $E$ :
\beq \label{ELOSS}
-\,\frac{d E}{ d z}\,\,\propto\,\frac{N_c \as}{ 2 \pi}\,\cdot\,\frac{1}{
\frac{1}{2} \,+\,\frac{\o_0}{4}} \,\cdot\,E^{\frac{1}{2} \,+\,\frac{\o_0}{4}}\,
\,.
\eeq
However, we shall be very careful with such kind of estimates since the
value of $\o_0$ depends crussially on the value of the gluon virtuality which
depends  on $\o$ in its turn (see \eq{ANSWERG}). The integral over $\o$
in $\frac{d E}{d z}$ concentrates at $\o \,\approx\,\,E$ where the soft energy
 approximation is not valid. Because of this we can consider \eq{ELOSS}
only as a rough estimate which we are going to improve later.
\section{Conclusions}
In this letter we have considered the LPM effect for photon and gluon
radiation off a fast quark propagating in a nuclear matter. The close
analogy betwen photon and gluon emission suggested in ref. \cite{BDPS}
has been confirmed and the relation between the Mueller approach \cite{MUT}
and traditional calculations has been established.
The main result reads:
\beq \label{FINANS}
\o\,\frac{d I}{ d \o d z}\,\,=\,\,
\,\,\frac{32  \alpha}{9\,\pi}\,\,\sqrt{\frac{ \,\frac{\as \pi^2}{3}
\,\,( x\,G(x,  \frac{1}{r^2_{soft}\,\,\sqrt{\kappa}})\,\rho}{\tau}}
\eeq
where $\kappa$ is defined in \eq{AVERU} and
 for QED:
 $\alpha = \alpha_{em}\,; \frac{1}{\tau}\,=\,\frac{\o}{2 E^2}\,;
x\,\,=\,\,\frac{\VEV{p^2_{\perp}}}{m E}$.
For QCD:
$\alpha = \frac{N_c \alpha_{S}}{2}\,; \frac{1}{\tau}\,=\,\frac{1}{2\o }
\,;
x\,\,=\,\,\frac{r^2_{soft}}{\o \sqrt{\kappa}}$.

The answer does not depend on the  nonperturbative QCD
scale. It depends on  the ratio
$\frac{\mu^2}{\lambda} \,=\,\rho\,\,\cdot\,\,\{ \,\frac{\as \pi^2}{3}
\,\,( x\,G(x,  \frac{1}{r^2_{soft}\,\,\sqrt{\kappa}}\,\}$ which
 depends on the QCD
 scale only in the argument of the gluon structure function. However, the value
of $\o$ for which we can apply the above formula comes from $\kappa\,\ll\,
r^2_{soft}$ and crucially depends on the value of $r^2_{soft}$ ( see \eq{OBHG})
{}.
$r_{soft}$ establishes the scale of distances
 ( $r_{\perp} \,\,<\,\,r_{soft}$ )
 where we can trust the perturbative  QCD approach  for nucleon interactions.
This scale has clear nonperturbative origin, however, it becames large and
grows with $x$ in the small $x$ region where the saturation of gluon density
in a nucleon  should be reached \cite{GLR}. Taking $r^2_{soft}\,\,
=\,\, 0.5  GeV^{-2}$
 and $\rho \,=\,0.17 Fm^{-3}$ we derive that our formula
 for gluon induced radiation can be justified for $ \o\,>\,\o_{BH} = 100 GeV$.
We are going to present more reliable estimates, as well as the application
 to the gluon inclusive cross section,  in further publications.

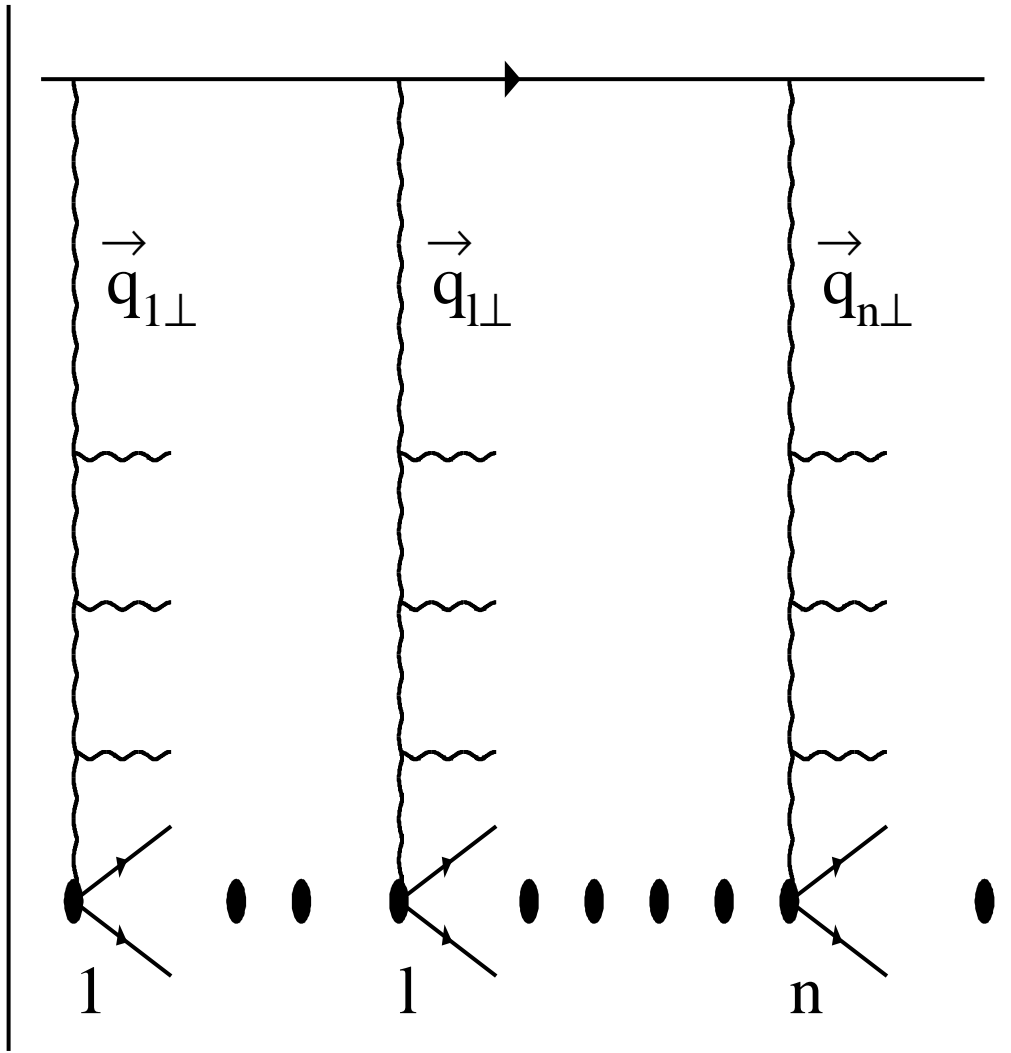
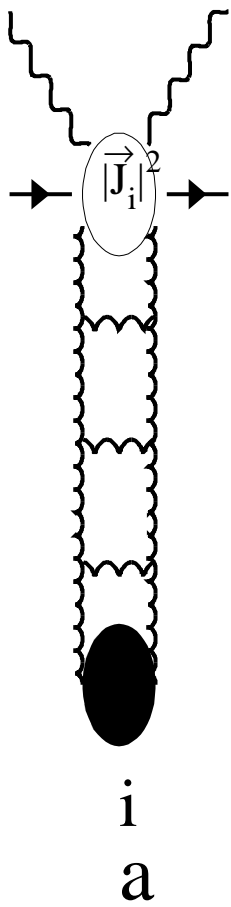
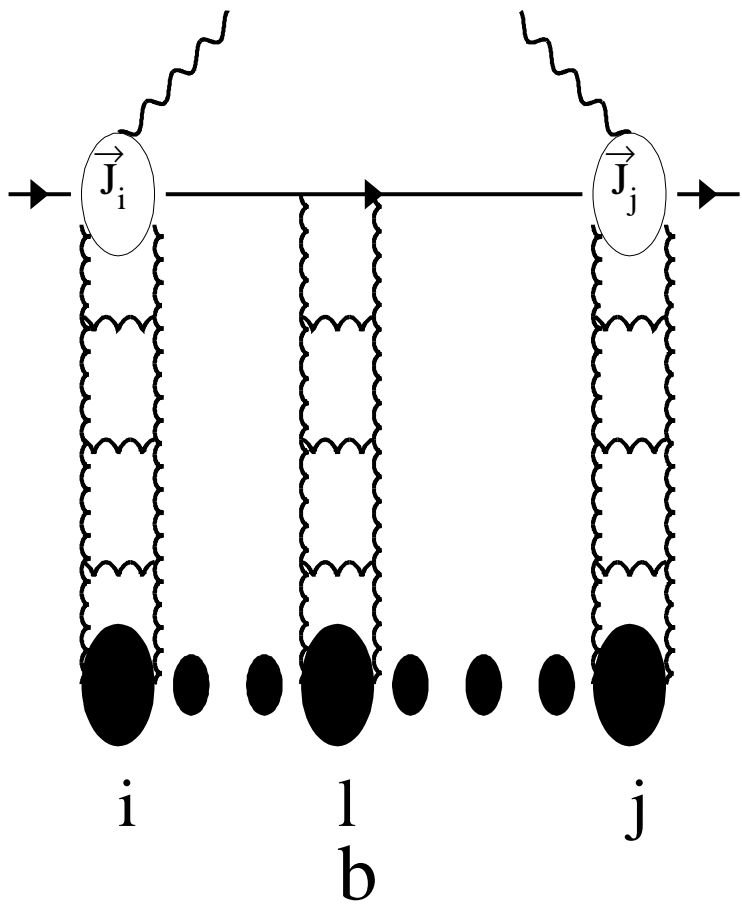
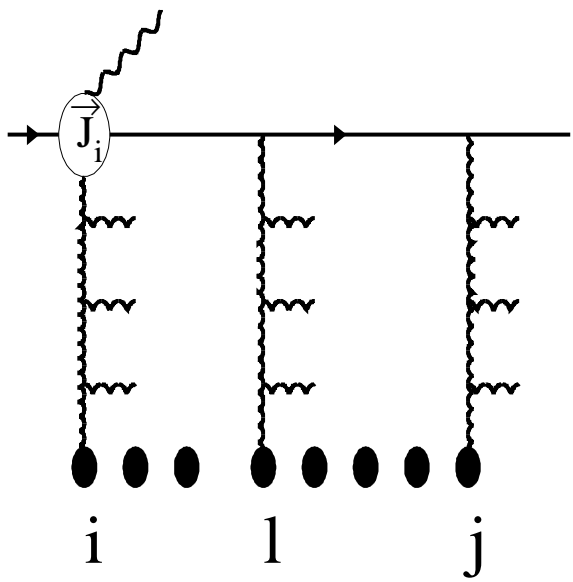
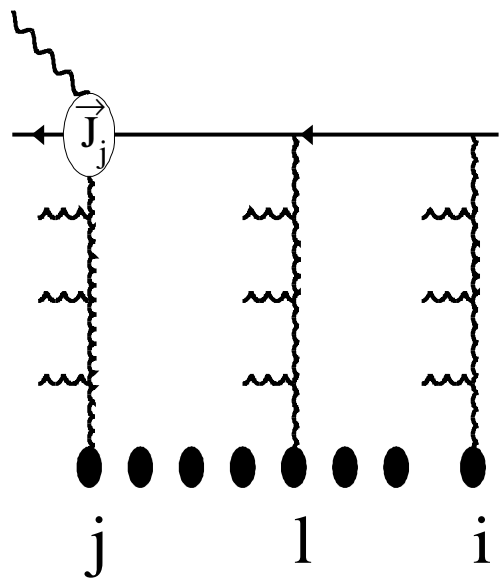
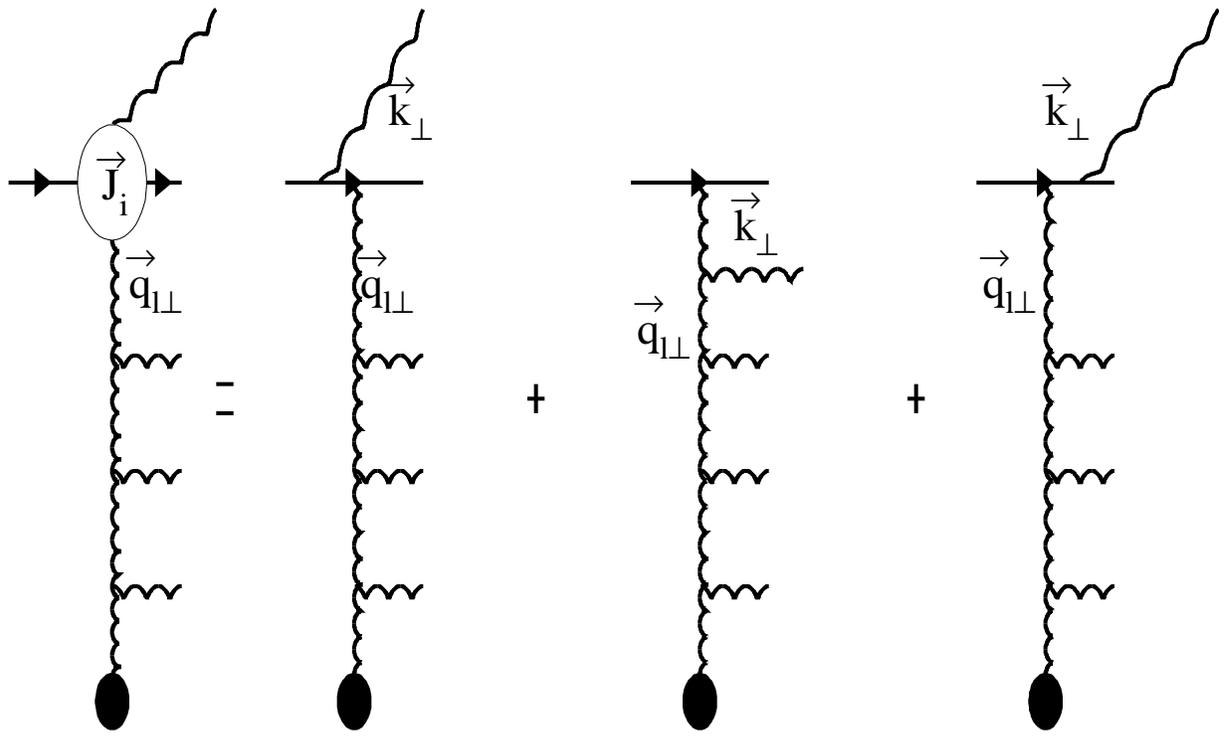
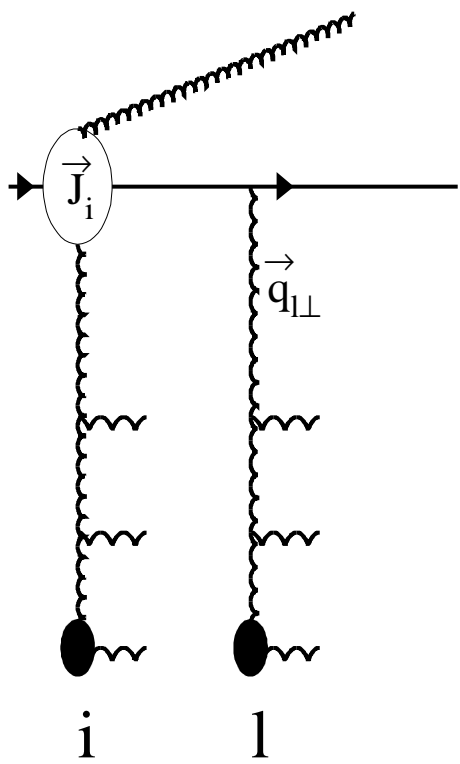
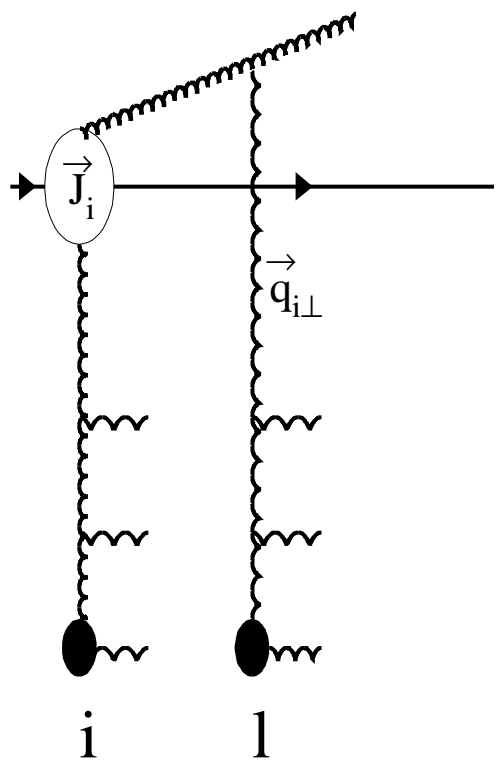
\end{document}